\documentclass{article}
\usepackage{ijcai22}

\usepackage{times}

\usepackage{soul}
\usepackage{url}
\usepackage[hidelinks]{hyperref}
\usepackage[utf8]{inputenc}
\usepackage[small]{caption}
\usepackage{graphicx}
\usepackage{amsmath}
\usepackage{booktabs}

\usepackage{amsfonts}
\usepackage{bm}
\usepackage{letltxmacro}
\usepackage{enumitem}
\usepackage{lipsum}  
\usepackage{subcaption}

\usepackage{booktabs}
\usepackage{multirow}
\usepackage[normalem]{ulem}
\useunder{\uline}{\ul}{}

\newcommand{\prop}{CLOZER}
\LetLtxMacro{\originaleqref}{\eqref}
\renewcommand{\eqref}{Eq.~\originaleqref}

\usepackage{bibunits}
\defaultbibliography{ijcai22}
\defaultbibliographystyle{named}
\newcommand{\citep}[1]{\cite{#1}}
\newcommand{\citet}[1]{\citeauthor{#1}~\shortcite{#1}}

\newcommand{\nibf}[1]{\noindent\textbf{#1}\ }

\title{Mask and Cloze: Automatic Open Cloze Question Generation\\
using a Masked Language Model
}

\author{
  Shoya Matsumori\and
  Kohei Okuoka\and
  Ryoichi Shibata\and\\
  Minami Inoue\and
  Yosuke Fukuchi\And
  Michita Imai
  \\
  \affiliations
  Keio University\\
  \emails
  \{shoya, michita\}@ailab.ics.keio.ac.jp
}

\begin{document}

\twocolumn[{%
\renewcommand\twocolumn[1][]{#1}%
\maketitle
\begin{center}
    \centering
    \captionsetup{type=figure}
  \includegraphics[clip,width=\linewidth]{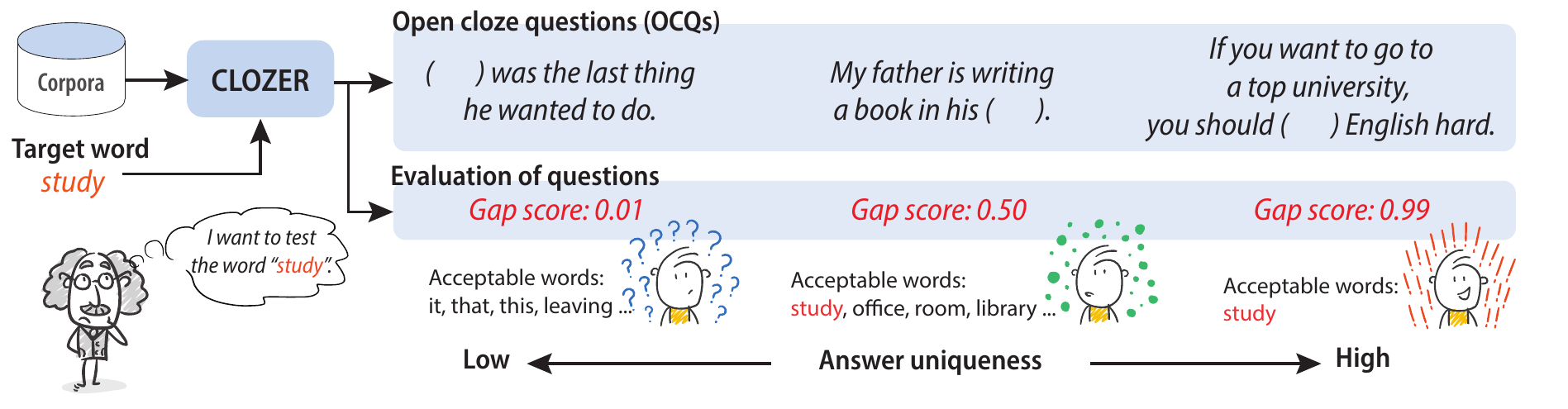}
  \vspace{-2em}
    \captionof{figure}{\textbf{
  	Given a target word that a student wants to learn or a teacher wants them to learn, CLOZER generates suitable OCQs for vocabulary learning from arbitrary corpora on the basis of a novel metric called a gap score. A question with a higher gap score only accepts the target word and does not allow other words in the blank; therefore, it represents the typical usage of the word.
    }}\label{fig:eyecatch}
\end{center}%
\vspace{1em}
}]


\begin{abstract}
Open cloze questions have been attracting attention for both measuring the ability and facilitating the learning of L2 English learners. In spite of its benefits, the open cloze test has been introduced only sporadically on the educational front, largely because it is burdensome for teachers to manually create the questions. Unlike the more commonly used multiple choice questions (MCQ), open cloze questions are in free form and thus teachers have to ensure that only a ground truth answer and no additional words will be accepted in the blank. To help ease this burden, we developed CLOZER, an automatic open cloze question generator. In this work, we evaluate CLOZER through quantitative experiments on 1,600 answers and show statistically that it can successfully generate open cloze questions that only accept the ground truth answer. A comparative experiment with human-generated questions also reveals that CLOZER can generate OCQs better than the average non-native English teacher.
Additionally, we conduct a field study at a local high school to clarify the benefits and hurdles when introducing CLOZER. The results demonstrate that while students found the application useful for their language learning. Finally, on the basis of our findings, we proposed several design improvements.
\end{abstract}

\section{Introduction}
\vspace{-0.2em}
Answer a word that will fit in the blank in the following sentence: \textit{``If you want to go to a top university, you should (~~~~) English hard.''}\footnote{The answer is \textit{study}.} Some of you might have struggled to answer such a question in the past. A question like this that asks you to fill in a gap with a word is called an Open Cloze Test or Open Cloze Question~(OCQ)~\cite{taylor1953cloze}, and it is widely used in language assessment tests for second language~(L2) learners, such as in Cambridge Assessment English tests. Compared to the commonly used Multiple Choice Question~(MCQ), where both the correct answer and several wrong answers (often called detractors) are provided for each question, the OCQ provides fewer clues for coming up with the right answer, and thus is considered to require higher cognitive abilities than simple reading skills~\cite{raymond1988close,mizumoto2016comparison}. OCQs are therefore useful not only for assessing the learner's abilities over a wide range but also for prompting language acquisition~\cite{taylor1957cloze,bormuth1967comparable,oller1979language}.

Introducing the OCQ to the educational field has many benefits. However, manual creation of the OCQs requires the designers to consider multiple aspects of a question, which is often overly cumbersome and time-consuming for teachers. Especially, to make the OCQs educationally beneficial, teachers have to ensure that only a ground truth answer and no additional words will be accepted in the blank~\cite{pino2008selection}.
A promising approach to such problems is automated question generation. However, previous studies on automated question generation have focused almost entirely on MCQs~(e.g., \cite{sumita2005measuring,mitkov2003computer,brown2005automatic}), and despite its educational importance, research on the generation of OCQs is relatively rare. Hence, it remains unclear whether algorithms can generate suitable OCQs and what the hurdles might be for introducing such methods to the educational field.

In this work, we propose \prop{}, an automatic OCQ generation algorithm~(Fig.~\ref{fig:eyecatch}). Given a target word that a student wants to learn or a teacher wants them to learn, \prop{} can generate unique answer OCQs that only accept the target word from arbitrary corpora. Questions that only accept the target word are expected to include the typical usage of the target word. As such, utilizing such questions will help students to grasp the actual concept of words. \prop{} generates questions by prediction and assessment: first, it predicts words that are likely to fit in a blank, and then, on the basis of the prediction, it assesses the answer uniqueness of a question. A masked language model~(MLM)~\cite{devlin2019bert} is utilized for the prediction. Given a sentence with a blank, MLM will predict a set of candidate words and assign them confidence values indicating the feasibility of each word belonging in the blank. However, simply using the ranking of the confidence values is not sufficient to guarantee the answer uniqueness. To this end, we introduce a novel metric called a gap score that measures how likely it is the target word will be the only answer. Our design of the gap score was inspired by the Gini coefficient, a metric used in economics to show wealth or income inequality. With the gap score, we can ensure not only the feasibility of the target word to the blank but also the exclusivity of the other words to the blank.

We performed a qualitative experiment by collecting answers from native English speakers and found that the gap score reliably represents the answer uniqueness of a question. Additionally, we conducted a field study at a local high school and proposed several design improvements based on the results.

In summary, our contributions are fourfold:
\begin{itemize}[topsep=0.5pt, itemsep=0pt]
\item We proposed an automatic OCQ generation algorithm. While the generation of MCQs has been studied extensively, there is much less research on OCQs. The findings in this work will make a valuable contribution to the field of education.
\item We pioneered the use of MLM for the automatic generation of OCQs. MLM has a wide range of applications in NLP, but none of the previous studies have focused on utilizing its inference ability to generate OCQs.
\item We introduced the gap score to measure the feasibility of a question by ensuring that the target word would uniquely fit into the gap. Quantitative experiments showed that the gap scores could accurately represent the uniqueness of a question.
\item We conducted a proof of concept through a field study and clarified the possible benefits and hurdles when introducing \prop{} to the educational field. 

\end{itemize}

\section{Related Works}
\vspace{-0.7em}
\nibf{Automatic generation of MCQs.}
Most of the prior works on the automatic question generation have focused on MCQs ~\cite{sumita2005measuring,mitkov2003computer,brown2005automatic,lee2007automatic,lin2007automatic,sakaguchi2013discriminative,goto2010automatic}. In general, there are three main steps required for generating MCQs: (1) selecting sentences from arbitrary sources, (2) determining a blank part from each sentence, and (3) generating distractors for each blank. Among such procedures, methods for (3) generating distractors have been intensively studied since distractors are considered to have a significant influence on the quality of a question~\cite{kurdi2020systematic}.
%
Conversely, (1) sentence selection and (2) determining a blank part have not been studied as much.
Sentences are usually chosen from arbitrary sources, such as a corpus or a textbook.
\citet{majumder2015system} proposed a sentence selection method that uses topic modeling and parse structure similarity. Most prior studies have utilized a fixed strategy for the gap selection; for example, \citet{sumita2005measuring} selected the leftmost single verb, and \citet{lin2007automatic} only selected an adjective as a blank. One of the few exceptions is a work by \citet{goto2010automatic}, which proposed a selection method based on conditional random fields (CRFs).

\nibf{Automatic generation of OCQs.}
While the automatic generation of MCQs has been studied extensively, there is much less research on OCQs~\cite{kurdi2020systematic}. One of the earliest studies was done by \citet{becker2012mind}, who addressed the problem of evaluating the quality of OCQs in a data-driven manner. Their  \textit{Mind the Gap} dataset was collected from 85 crowdsourcing workers and includes annotations for the quality of questions. Additionally, \citet{malafeev2014language} developed an automatic OCQ generation algorithm that aimed to emulate the OCQs found in Cambridge Assessment English tests. Similar to our approach, \citet{felice2019entropy} assumed that the fewer possible words for the gap, the easier the question, and introduced information entropy~\cite{shannon1948mathematical} to measure the complexity of a gap.



\nibf{Limitations of previous studies.}
The limitations of these previous studies are summed up as follows. 
First, the gap generation for MCQs~\cite{sumita2005measuring,lin2007automatic,goto2010automatic} is not applicable to OCQs because there is no demand to ensure uniqueness in MCQs and it is not designed to do so.
While prior studies have addressed the generation of gaps in OCQs, several hurdles still exist.
The n-grams utilized in \citet{felice2019entropy} cannot reflect longer context due to the computational costs.
In fact, only a 5-gram model was used in that study, which is problematic if we want to consider a longer context that requires referencing words more than four words distant from the target word.
As for other studies~\cite{malafeev2014language,becker2012mind}, they require a dataset to train the model but collecting such a dataset imposes substantial costs.
Additionally, their interests were to generate OCQs from arbitrary text sources and they cannot be applied to generation based on the target words.
Finally, the above studies focused only on evaluating their algorithms, and the design requirements for the application are still unknown.

\begin{figure*}[th]
\centering
  \includegraphics[clip,width=1.0\linewidth]{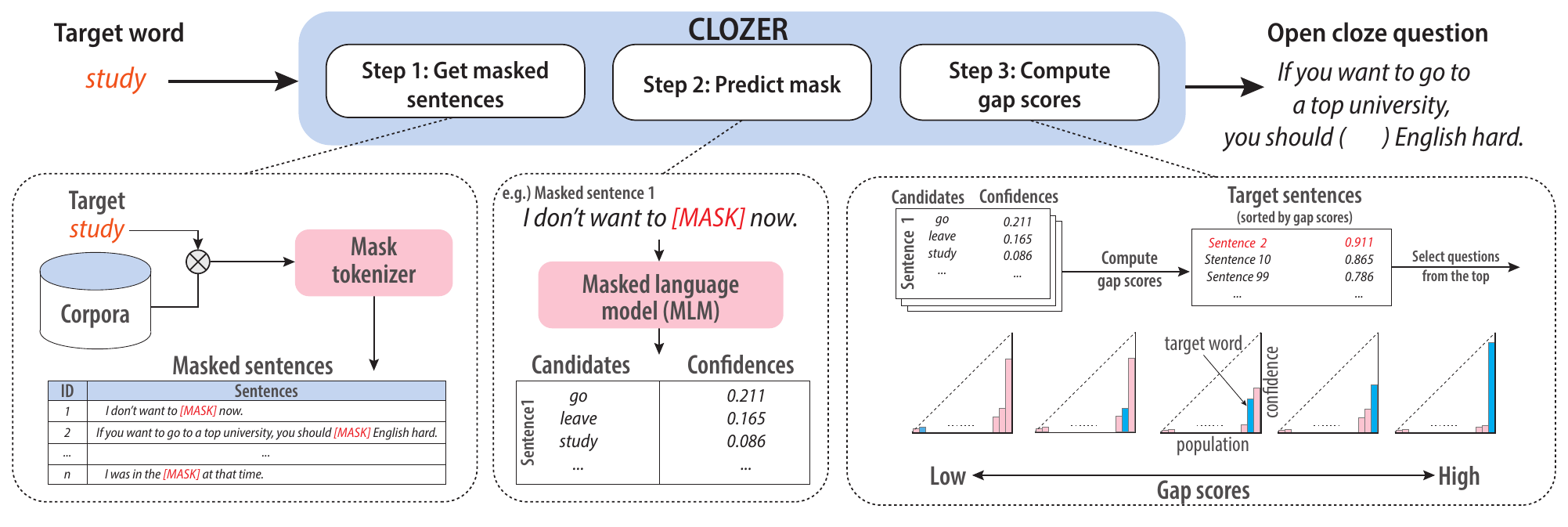}
  	\vspace{-8mm}
  	\caption{The three generation steps of CLOZER. (Step 1) Given a target word, target sentences are extracted from arbitrary corpora and then are converted into masked sentences, whose target words are substituted with a mask token. (Step 2) Given a set of masked target sentences, a masked language model predicts words and confidences for the mask token. (Step 3) On the basis of the mask prediction, a gap score for each target sentence is computed, and sentences with high gap scores are selected as questions.
    }
  	\label{fig:model}
\end{figure*}
\section{Methods and Implementations}
The general flow of \prop{}, an automatic open cloze question generator for computer-mediated language learning, is shown in Fig.~\ref{fig:model}. Given a target word that one desires to learn, \prop{} generates open cloze questions based on a gap score. The learning goal of CLOZER is to improve vocabulary skills. A sentence whose gap score is high means that the target word is likely to be the only answer to the blank. In other words, such a sentence is expected to embody the typical usage of the target word. By asking questions whose gap scores are high, students can learn the most specific use of the target words. The generation process can be divided into three steps: fetching masked target sentences, predicting masks, and computing gap scores.

\subsection{Step 1: Fetch masked target sentences}
Given a target word $w$, a set of sentences $\mathcal{S}_w$ that includes $w$ is collected from an arbitrary corpus $\mathcal{S}$ as
\begin{align}
    \mathcal{S}_{w} = \{s \ | \ s \in \mathcal{S} \wedge w \in s\}.
\end{align}
Since $\mathcal{S}_w$ may include some words that are unfamiliar or improper for learners, they  are filtered out in accordance with a pre-defined word list $\mathcal{W}$. Sentences that are too short or too long are pruned as well. The target sentences ${\mathcal{S}}_w$ are then converted into masked target sentences $\widetilde{\mathcal{S}}_w$, where the target word is substituted by a special mask token.

\subsection{Step 2: Predict masks}
Given a masked target sentence $\tilde{s} \in \widetilde{\mathcal{S}}_w$, the masked language model $F_{\text{M}}$ yields the tuple of a set of candidate words $\lambda$ along with corresponding confidence values $\bm{c}$, as
\begin{align}
    (\lambda, \bm{c}) = F_{\text{M}}(\tilde{s}),\label{eq:mask}
\end{align}
where $\lambda = \{w_i \ | \ i \in \mathcal{I}_\text{v}\}$ and $\bm{c} = [c_i \ | \ i \in \mathcal{I}_\text{v}]$, $\mathcal{I}_\text{v}$ is an index set whose size is the number of vocabularies, and both $\lambda$ and $\bm{c}$ are sorted in the descending order determined by $\bm{c}$.
The confidence score of the masked prediction $c_i$ represents the feasibility of a word $w_i$ being in the position of the masked token.

\subsection{Step 3: Compute gap scores}
On the basis of the tuple $(\lambda, \bm{c})$, the gap score $\varphi$ is computed. $\varphi$ represents the answer uniqueness of a question, namely, how uniquely the target word fits the gap.
Our design for the uniqueness is based on the following two criteria:
{
\setlength{\leftmargini}{1.5em}
\begin{itemize}
\item[1.] The target word must have high confidence value
\item[2.] The target word must have a significantly higher confidence compared to the other candidates
\end{itemize}
}
A question is required not only to be in a higher confidence but also to have a significantly higher confidence compared to other word candidates. This is because there are cases where the target word has high confidence, for example, 0.8, but the second possible candidate has a lower but not significantly low confidence value, for example, 0.19. In such a case, it is possible that the second-best candidate will be chosen.

To meet the criteria, we borrow the idea of the Gini coefficient~\cite{giugni1912variabilita,dorfman1979formula}, a widely used metric in the field of economics that shows a wealth or income inequality existing between two specific groups. In a similar manner, we regard the confidence value as wealth or income and use the Gini coefficient to measure any inequality that exists between the confidence of the target words and that of the other candidates. Originally, the Gini coefficient is defined as 
\begin{align}
    G = 1 - 2\int_0^1 L(x)\ dx, \label{eq:gini}
\end{align}
where $L(x)$ is the Lorenz curve~\cite{lorenz1905methods}, which is a monotonic increasing function that takes the cumulative portion of the population $x$ and returns the cumulative portion of the total wealth or income. Since $L(x)$ is a convex function, the range of the Gini coefficient is $0 \leq G < 1$, where $G=0$ implies a state of complete equality and $G=1$ a state of complete inequality.

Given a set of confidences  $\bm{c}$, the Lorenz curve for the confidence values $\bm{c}$ is defined as a discrete function:
\begin{align}
    L(\bm{c}, i) = \frac{\sum_{j=1}^{i}\bm{c}_j}{\sum_{j=1}^{N_{c}}\bm{c}_j},
\end{align}
where $\bm{c}$ is sorted in descending order and $N_c$ is a size of $\bm{c}$. The Gini coefficient for $L(\bm{c}, i)$ is derived as a discrete form of \eqref{eq:gini}, as
\begin{align}
    f_{\text{gini}}(\bm{c})
    = 1 - 2\sum_{i=1}^{N_c}\frac{1}{N_c}\frac{L(\bm{c}, i)+L(\bm{c}, i-1)}{2},
\end{align}
whose range is $0 \leq f_{\text{gini}}({\bm{c}}) < 1$. Akin to the original interpretation, $f_{\text{gini}}(\bm{c}) = 1$ represents that the confidence values are monopolized by the top-1 word and $f_{\text{gini}}(\bm{c}) = 0$ represents that they are fairly distributed to each word.

Accompanied by $f_{\text{gini}}$, the equation for the gap score is defined as
\begin{align}
    \varphi(\bm{c}, j) = f_{\text{gini}}(\bm{c}_{j:}) \cdot f_{\text{rw}}(\bm{c}, j) 
    ,\label{eq:gap}
\end{align}
where $j$ is the index of the target word in $\lambda$, \textit{i.e.}, $w_j = w$.
$f_{\text{rw}}$ is a reweighting function introduced to shrink the gap score based on the proportion of the confidence value of the target word and top-k confidences, as
\begin{align}
    f_{\text{rw}}(\bm{c}, j) = \frac{c_j}{\sum_{i=1}^{k}c_i},
\end{align}
where $k$ is a constant value.

Finally, by applying \eqref{eq:mask} and \eqref{eq:gap} for every masked sentence in $\widetilde{\mathcal{S}}_w$, we obtain the sorted gap scores for all sentences as
\begin{align}
    \Phi = \{\varphi_i \ | \ i \in \mathcal{I}_\mathcal{S}\},    
\end{align}
where $\mathcal{I}_\mathcal{S}$ is an index set whose size is the corpus size and $\varphi_i$ represents the confidence score for $i$-th $\tilde{s}$, which satisfies $\varphi_i \leq \varphi_{i+1}$.

\section{Quantitative Experiments}

Our research questions are:
\begin{itemize}[topsep=0.5pt, itemsep=0pt]
    \item Whether the gap score accurately represents the uniqueness of a question.
    \item Whether and how the different MLM architectures affect the performance.
    \item How well CLOZER can generate unique questions compared to human experts.
\end{itemize}
In the experiment, we asked native English speakers to take a cloze test generated by CLOZER and compared the correlation coefficient between the gap score and correct ratio across several MLM variants.  Our expectation was that a question with a higher gap score will collect answers that match the target word, while a question with a lower gap score will collect a wide range of answers that are both syntactically and semantically feasible but not necessary the target word.
Additionally, we asked non-native high school English teachers to manually create OCQs and then compared them with the ones generated by CLOZER.

\subsection{Settings}
The questions were generated by choosing target words from the vocabulary list of EIKEN, a widely used English qualification exam in Japan.
The vocabularies were collected from Grade 5 to Grade 2, respectively equivalent to the first year of junior high school and to high school graduates.
Since CLOZER mainly focuses on vocabulary learning, only content words were chosen for the target words; function words (e.g., determiners, pronouns, prepositions, and conjunctions) were excluded.

The target sentences were collected from the Corpus of Contemporary American English~\cite{davies2009385+}, which includes 1.1 billion words collected from sources of various genres spanning spoken word performances, fiction, popular magazines, newspapers, academic texts, TV and movie subtitles, blogs, and other web pages. To compare the performance across different MLM architectures, we sampled 40 questions randomly from the target sentences after checking to make sure the questions did not include any inappropriate words or personal information. We utilized  eight model variants based on BERT~\cite{devlin2019bert}, DistilBERT~\cite{sanh2019distilbert}, RoBERTa~\cite{liu2019roberta}, and ALBERT~\cite{lan2019albert} that were trained  with masked language modeling\footnote{All of the pre-trained models were provided by Hugging Face~\cite{wolf2019huggingface}.}: bert-base-uncased, bert-large-uncased, distilbert-base-uncased, roberta-base-v2, roberta-large-v2, albert-base, albert-large, and albert-xlarge. 

We recruited participants from Amazon Mechanical Turk, the popular crowdsourcing platform. All participants were asked to read the instructions carefully and to answer 40 questions. Further details of the settings and the supplemental results can be found in the appendix.

\begin{table}[t]
\small
\centering
\begin{tabular}{@{}lcccc@{}}
\toprule
Model name       & $r\uparrow{}$                  & $p\downarrow{}$      & Parameters & Dataset \\ \midrule
bert-base        & 0.760                & 3.03E-16 & 108M       & 16GB    \\
bert-large       & \underline{0.791}    & 2.47E-18 & 334M       & 16GB    \\
distil-bert-base & 0.647                & 9.31E-11 & 66M        & 16GB    \\
roberta-base     & \textbf{0.800}       & 5.55E-19 & 123M       & 160GB   \\
roberta-large    & 0.745                & 2.28E-15 & 354M       & 160GB   \\
albert-base      & 0.524                & 6.03E-07 & 12M        & 16GB    \\
albert-large     & 0.671                & 9.76E-12 & 18M        & 16GB    \\
albert-xlarge    & 0.671                & 9.40E-12 & 235M       & 16GB    \\ \bottomrule
\end{tabular}
\caption{Pearson's correlation coefficients ($r$), p-values ($p$), number of parameters, and size of the dataset used in pre-training. All models were statistically significant~($p < 0.001$). 
}\label{srt:model_comp}
\vspace{-1.5em}
\end{table}

\subsection{Statistical results}
One thousand six hundred answers were collected from 40 workers, all of whom were native English speakers. 
The duration of each task was $41.24{\pm 28.11}$ minutes on average. The average correct ratio (i.e., the ratio of matches between the target word and the submitted answer) was $40.58{\pm 9.33}\%$, where the highest score was $55.00\%$ and the lowest was $20.00\%$.

Table~\ref{srt:model_comp} indicates that the gap score is effective as a metric for measuring the uniqueness of OCQs.  Specifically, roberta-base performed the best and bert-large-uncased was the second best, with the correlation coefficients of 0.800 and 0.791, respectively. In contrast, the performance of the DistilBERT and ALBERT variants was lower than the BERT and RoBERTa variants. This is presumably due to the side effect of lightening the model size, considering both DistilBERT and ALBERT have made efforts to shrink the number of parameters. Although increasing the parameter size in the same architecture generally improved the correlation coefficient, this was not the case with RoBERTa. In roberta-large, some questions yielded higher gap scores than expected. In other words, the model was able to predict true answers better than humans, and this caused the correlation coefficient to degrade.

\begin{figure}[t]
  \centering
  \includegraphics[width=\linewidth]{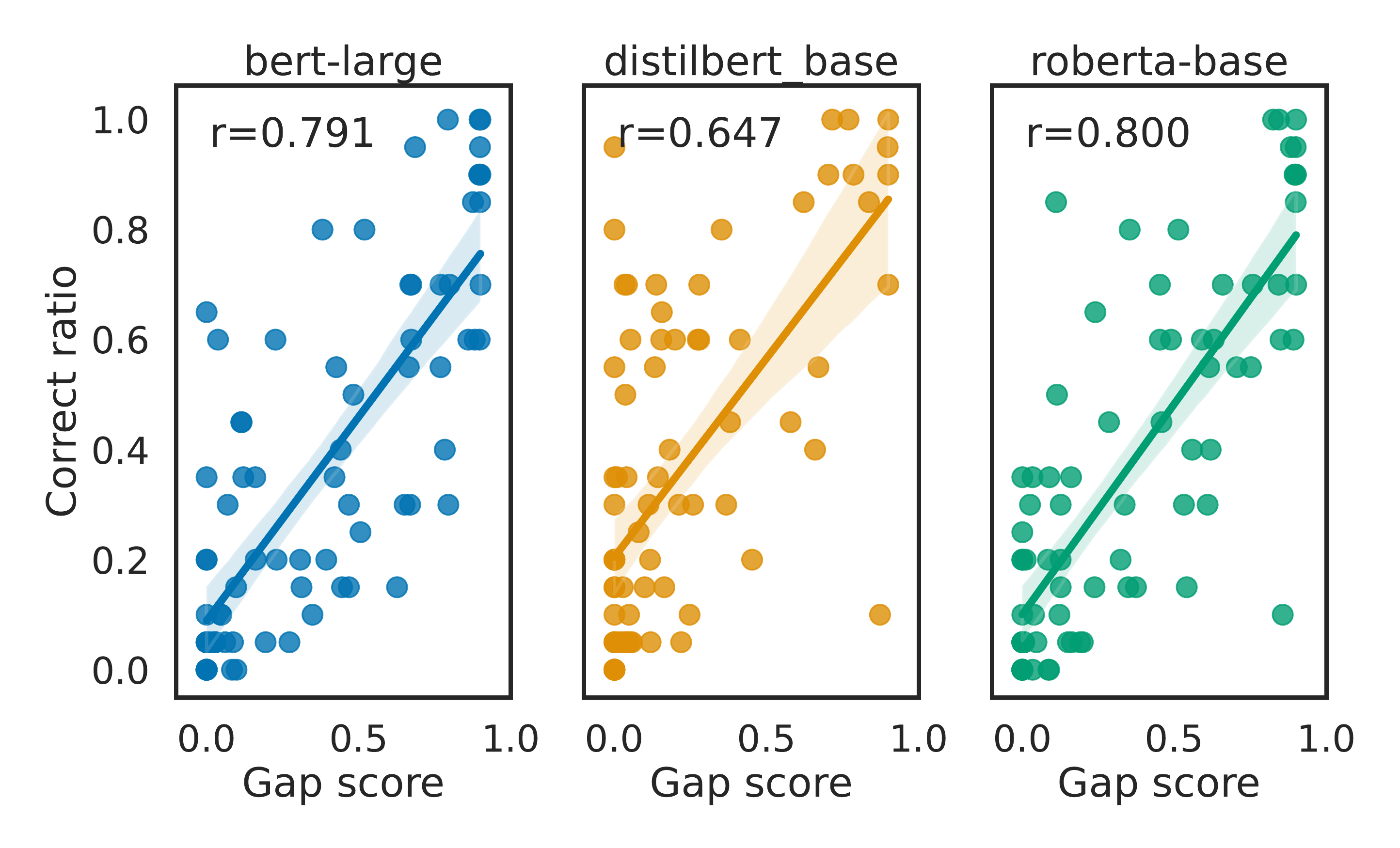}
  \vspace{-2em}
  \caption{
  Comparison of correlation plots between correct ratio and gap score.
  }
  \label{fig:gap_vs_corr}
 \vspace{-1em} 
\end{figure}

Figure ~\ref{fig:gap_vs_corr} shows the scatter plots of the gap score and correct ratio. In the highly correlated models, bert-large and roberta-base, a higher gap score indicates a higher correct score. There were comparably few samples that had high gap scores but low correct ratios. Some of the samples with lower gap scores had higher correct ratios, but this is acceptable if their number is not too large, since the users will usually choose questions whose gap scores are high and ignore the other samples. On the other hand, the model with lower correlation, distil-bert, tended to under-rate samples with a high correct ratio.

\subsection{Representative samples}
We further investigate the four representative cases obtained by the best-performed model, roberta-base: cases where the gap score and the correct ratio were both high or both low, and cases where one of the two was high.

A sample for the first case, where both the gap score and the correct ratio were high, is ``Just for my own (\ \ \ \ ) of mind, Ted, send me a list of your exes' names and current addresses.'' The target word for this question was \textit{peace}. The gap score for this target sentence was $0.90$, and the correct ratio was $90\%$. This type of question will only accept the target answer and, therefore, will be selected as a cloze question.

In contrast, a sample where both the gap score and the correct ratio were low is ``She impressed me as a rather (\ \ \ \ ) but pleasant woman who initially seemed quite timid and shy.'' The target word was \textit{serious}, and the gap score and the correct ratio were both $0.0$. A question where the gap score is low is expected to accept various words other than the target word, and thus, will not be selected as a question.

An example of the case where the gap score was high but the correct ratio was relatively low is ``There is a point, when trying to (\ \ \ \ ) a turn on a motorcycle, when you know you cannot negotiate the turn.'' The target word for this sentence was \textit{negotiate}, and the gap score and the correct ratio were respectively 0.86 and 10\%.
Interestingly, the typical answer was \textit{make}, which is also acceptable. We assume most workers made early decisions without noticing the same phrase ``negotiate the turn'' in the following clause. This suggests that the model tended to be too confident when the target word was repeated in a similar phrase. Filtering out such cases may prevent this problem.

The final sample is the case where the gap score was low but the correct ratio was high. An example is ``Not long after that, however, they'd discovered the caves lying beneath those mountains, and the species which should have gone (\ \ \ \ ) lived on.'' The target word for this sentence was \textit{extinct}. While the correct ratio of the question was 85\%, the gap score was 0.11 and the top-1 prediction of the model was \textit{underground}. Again, this case is more acceptable than the third case since these questions are discarded and not selected as questions in the end.

\begin{table}[t]
\centering
\small
\begin{tabular}{@{}lcc@{}}
\toprule
Method      & \multicolumn{1}{c}{Correct ratio (\%)} & \multicolumn{1}{c}{Gap score} \\ \midrule
Human~(avg)       & $53.4\pm34.7$ & $0.367\pm0.348$                         \\
Human~(best)       & $88.0\pm32.6$ & $0.667\pm0.211$                         \\
CLOZER$_{0.8}$ roberta-base & $79.6\pm24.6$ & $0.876\pm0.013$\\
CLOZER$_{0.8}$ bert-large & $82.1\pm15.4$  & $0.891\pm0.027$\\
\bottomrule
\end{tabular}
\caption{Comparisons between human-generated and CLOZER-generated questions. (avg) represents the average performance and (best) represents the best performance. CLOZER$_{0.8}$ represents CLOZER-generated questions whose gap scores were higher than 0.8. 
}\label{srt:human_gen}
\vspace{-1.5em}
\end{table}
\subsection{Human-generated questions}
Additionally, we conducted an experiment to compare human-generated questions and auto-generated questions. Since we expect most of the teachers using CLOZER to be non-native English speakers, we recruited non-native high school English teachers at a high school in Japan. Eight teachers (average age: $46.12\pm{}11.87$) participated in the experiment to collect 40 questions in total. In the experiment, teachers were asked to create questions based on the same designated target words that were used to generate questions in the previous experiment. Teachers were allowed to consult any materials during the experiment unless they directly copied or modified existing OCQs. It took $19.13\pm{}8.04$ minutes on average to create five questions.

The questions were then used to collect 1,600 answers from 40 native English speakers at Amazon Mechanical Turk. The results are presented in Table~\ref{srt:human_gen}, where Human (avg) represents the average results for human-generated questions, Human (best) represents the results for questions from the best-performing participant, and CLOZER$_{0.8}$ represents the results for CLOZER-generated questions whose gap scores were higher than 0.80. The correct ratio and gap score for the Human (avg) condition were comparably lower than those of the CLOZER variants. The best-performing result for the human condition was $88.0\%$, which is better than the CLOZER variants, although its deviation was higher than the CLOZER conditions. These results indicate that CLOZER can generate higher-quality single answer questions than average non-native English teachers.

\section{Field Study}
We conducted a field study at a local high school to clarify the benefits and hurdles when introducing open cloze tests generated by CLOZER to the educational front. For this purpose, a prototype of the CLOZER application, ``CLOZER proto'', was deployed on a web platform accessible from any device using a browser.

\subsection{Settings}
The field study was conducted at a high school in Kanagawa, Japan. We recruited 24 high school students in either their first or second year. The study was conducted for two days with different participants: 14 and 10 students on the first and second days, respectively. On the second day, a hint was added to each question to ease the difficulty. The hint was the first letter of the ground truth word given on each question when participants had answered incorrectly on their first try. The students were asked to perform a cloze test consisting of 20 OCQs generated by CLOZER, to fill in a questionnaire, and to take an interview. 
We followed the same strategy as the quantitative experiments to generate target questions.
Specifically, we set the level of vocabulary used in the questions so that it would be familiar to high school students by restricting the target words and target sentences with the word list.
To determine the upper bound of the correct ratio for the questions, we also collected answers from 20 native English speakers using Amazon Mechanical Turk. The conditions for the task were the same as in the quantitative experiments, except for the number of questions. Further details of the settings and the supplemental results can be found in the appendix.

\begin{table}[t]
\scriptsize
\centering
\begin{tabular}{@{}lcccc@{}}
\toprule
 & \multicolumn{2}{c}{Average (\%)} & \multicolumn{2}{c}{Best (\%)} \\ 
Participants & Exact & Stem & Exact & Stem \\ \midrule
Students (day 1) & $11.8\pm11.2$ & $13.2\pm12.7$ & 35.0 & 45.0 \\
Students (day 2 w/o h) & $8.0\pm8.9$ & $8.5\pm9.1$ & 25.0 & 25.0 \\
Students (day 2 w/ h) & $19.5\pm8.3$ & $23.0\pm9.5$ & 35.0 & 35.0 \\
Native speakers & $84.0\pm9.7$ & $84.5\pm9.4$ & 95.0 & 95.0 \\ \bottomrule
\end{tabular}
\caption{Correct ratios in field study. ``Exact'' represents the exact match ratio and ``Stem'' represents the stem match ratio. Since hints were given on day 2, correct ratios without (w/o h) and with (w/ h) hint conditions are presented.}\label{srt:student_sum}
\vspace{-1.5em}
\end{table}

\subsection{Results}
Table~\ref{srt:student_sum} shows the results of the correct ratio for each group of participants. The correct overall ratio was measured by two metrics: exact match and stem match. The exact match regards the answer to be correct when it is identical to the ground truth, and the stem match is correct when the stem of the answer matches the stem of the ground truth. The results show that the average correct ratios for students were significantly  lower than those of native speakers. This was a little surprising, since we designed the level of vocabulary used in the questions to correlate with the students’ proficiency level. This indicates that, even with basic words that all students must learn, there is still a difference in the depth of vocabulary knowledge between non-native students and native English speakers.
Additionally, we found that the deviations of the correct ratios were high; some students managed to answer the question much better than the others and vice versa. This was also surprising, since the students had similar English levels. For example, 11 students out of 14 in day 1 were qualified with EIKEN Grade Pre-2, which is equivalent to the European Framework of Reference for Languages (CEFR) A2 level. This indicates that CLOZER revealed vocabulary skills that cannot be measured by the existing learning materials.

\subsection{Discussion}
While most of the students mentioned in their interviews that CLOZER was useful for language learning, alleviating the difficulty is necessary for the continual use of the application. Hence, we will discuss the features required for future design improvements as based on our findings in the interviews.

\nibf{Grammatical features.} A few students reported that the grammar was difficult. 
L2 English learners are generally dependent on grammar rules, unlike native speakers who operate such rules unconsciously. One solution would be to show a part of speech and a dependency of each word. The part of speech (such as noun, verb, and adjective) indicates how the word functions in terms of both meaning and grammar within the sentence, while the dependency shows which word is related to another word as well as the type of the relationship between them. With such information, users will be able to use their grammatical knowledge to understand the meaning of a sentence. 

\nibf{Extra context.} In the test, most of the students struggled to grasp the context of a sentence. This can be improved by adding some extra sentences before or after the target sentence. We only used a single sentence for each question in this study, and although the native speakers on MTurk managed to interpret contexts, one sentence was too short for L2 learners. Adding extra sentences will help them understand in what context the sentences were written. Showing the translation for such sentences will also be helpful.

\nibf{Hints for the target word.}
Lastly, some students understood the context but found it difficult to come up with the answer. In such a case, it will be helpful to add additional information to the target word: for example, to show the first letter. As we found on day 2 of the field study, this hint was effective for most students. We could also increase the number of letters shown to ease the difficulty further. Similarly, showing other types of information about the target word, such as the number of letters, would also help decrease the difficulty.

\section{Conclusion}
In this work, we presented CLOZER, an automatic open cloze question generator.
The results of quantitative experiments statistically demonstrated that it can generate open cloze questions that only accept the ground truth answer. 
Furthermore, the field study at a local high school revealed possible benefits of CLOZER.
Finally, on the basis of our findings, we proposed several design improvements.
Future research will focus on improving the functionality of the application and introducing CLOZER into educational settings to verify its effectiveness over the long term.


\clearpage
\appendix
\section*{APPENDIX}

\section{Ethical Considerations}
There has been quite a lot of work exploring potential bias in language models (e.g., \cite{kurita2019measuring}).
Since CLOZER utilizes a language model, there is a concern that some questions may contain biased word usage.
Several studies have proposed methods to mitigating bias~\cite{liu2021mitigating,huang2020reducing} and application of such procedures is necessary for practical use.
Additionally, teachers should be made aware of such risks so that they can devise ways to avoid selecting biased questions during the selection process.

\section{Supplemental Related Works}
\subsection{Cloze tests}
The cloze test (alternatively, open cloze test or cloze procedure) was invented by \citet{taylor1953cloze} in the early 1950s. The term \textit{cloze} stems from the Gestalt psychology theory of the principle of closure~\cite{koffka2013principles}, which mentions the human tendency to perceive complete patterns from partially hidden or incomplete patterns. Cloze tests can be easily created by deleting random words or every nth word from passages. While cloze tests were initially introduced as a measurement tool for the readability of prose passages, some researchers (including Taylor himself) later discovered that it was also applicable to the measurement of one's reading ability~\cite{taylor1957cloze,bormuth1967comparable}.

Research on cloze testing initially focused on native English speakers but was later applied to the field of L2 learning and teaching. Many studies revealed the relationships between cloze test scores and L2 learners' language skills and concluded that it was effective for measuring proficiency and therefore valuable as a placement test (e.g., \cite{friedman1974use,anderson1972application}). Variants of the cloze test, such as the C-test~\cite{klein1985cloze} and the rational cloze test~\cite{bachman1985performance}, which aimed to revise some of the weaknesses of the original cloze test, have also been accepted as measurement tools. There are also claims that cloze tests can be a good tool for measuring integrative language skills~\cite{oller1979language}. Although there is some debate regarding the extent to which cloze testing can actually measure language skills, its benefits are now generally accepted~(e.g. \cite{kobayashi2002method,weir1988communicative}).

Tests featuring multiple choice questions~(MCQ), which include both the correct answer and several wrong answers (often called distractors), have frequently been used as well~\cite{lado1961language}. However, with MCQ, test takers tend to find the answers simply by a process of elimination after inspecting the given choices, and thus it is considered to require limited language abilities~\cite{raymond1988close}. In fact, \citet{mizumoto2016comparison} reported that, compared to the cloze test condition, less brain activation was observed in the MCQ condition.

\subsection{The history of automatic question generation.}
With the development  of information technologies, massive open online courses (MOOCs) and other e-learning platforms  have gained attention recently~\cite{qayyum2018distance,gaebel2014moocs,hew2014students,bates2005technology}. In particular, in the field of computer-assisted language learning (CALL)~\cite{levy1996call}, researchers have examined various approaches for automatic question generation~\cite{wita2018semantic,killawala2018computational,afzal2014automatic,zavala2018use}. With the aid of automation, educators can reduce the cost of question construction, which frees them up to concentrate on more important activities. Additionally, having a large number of questions generated by a model enables additional teaching processes such as adaptive learning and repetitive drill practice to be utilized. Finally, a certain control of the question characteristics (e.g., difficulty) is useful for meeting the requirements of particular test settings.

\subsection{Language modeling}
Language modeling is used to obtain a language model that represents the probability distribution of words given certain context words. Language models have a wide range of applications, such as speech recognition~\cite{amodei2016deep}, machine translation~\cite{wu2016google}, summarization~\cite{rush2015neural}, text generation~\cite{brown2020language}, and text classification~\cite{howard2018universal}.

A typical language model is an n-gram~\cite{dunning1994statistical}. The n-gram refers to the number of $n$ sequential words that are used to predict a word. For example, the n-gram of $n=1$ is called a unigram, which only sees the current word, and $n=2$ and $n=3$ are called a bigram and a trigram, respectively, which see the current word and one or two history words. While the unigram can be considered as i.i.d., in the $2 \leq n$ cases, the n-gram can be considered as an $n-1$ order Markov process. 

Another approach for language modeling uses a neural network as an approximation function. Most of the studies~\cite{sundermeyer2012lstm} in this vein have used a recurrent neural network (RNN) for this purpose, such as LSTM~\cite{hochreiter1997long} or GRU~\cite{cho2014learning}. These approaches utilize left-to-right/right-to-left language modeling, which models the language sequentially by predicting the rightmost/leftmost token in the forward/backward direction or a combination of the two.

One of the most powerful language models currently is BERT~\cite{devlin2019bert}, which has been used in a novel language modeling task called masked language modeling~(MLM) for the Transformer architecture~\cite{vaswani2017attention}.
In MLM, similar to the cloze test, a certain portion of tokens (usually $15\%$) in the input sentences is randomly substituted with a special token called a mask token, and the model is trained to predict tokens for each of these masked tokens. Unlike the previous left-to-right or right-to-left language modeling, MLM has successfully obtained deep-bidirectional representations, demonstrating state-of-the-art performances in various downstream tasks~\cite{qiu2020pre}.
\begin{figure*}[t]
  \centering
  \includegraphics[width=\linewidth]{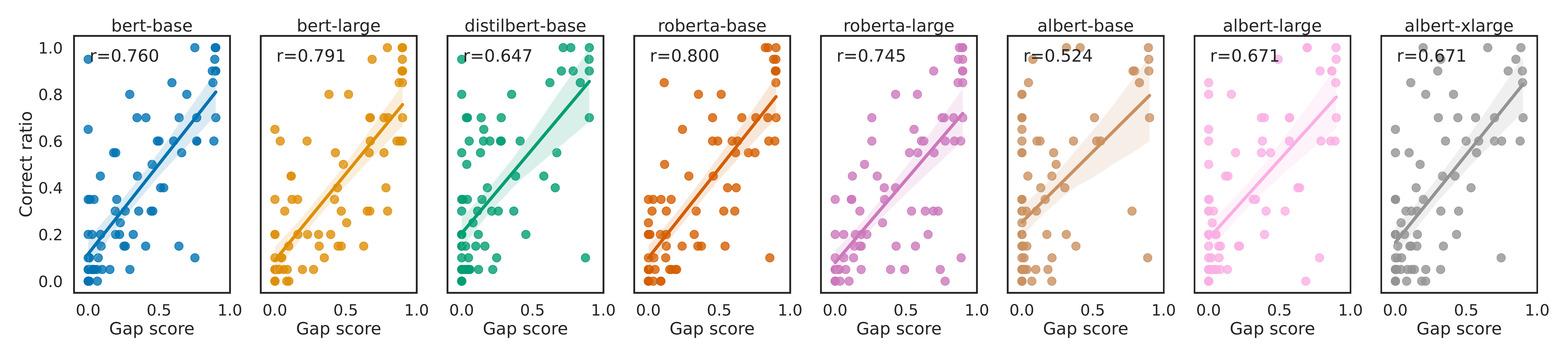}
  \caption{
  Comparison of correlation plots between correct ratio and gap score.
  }
  \label{fig:gap_vs_corr_all}
\end{figure*}
\begin{figure}[t!]
  \begin{center}
    \includegraphics[clip,width=\linewidth]{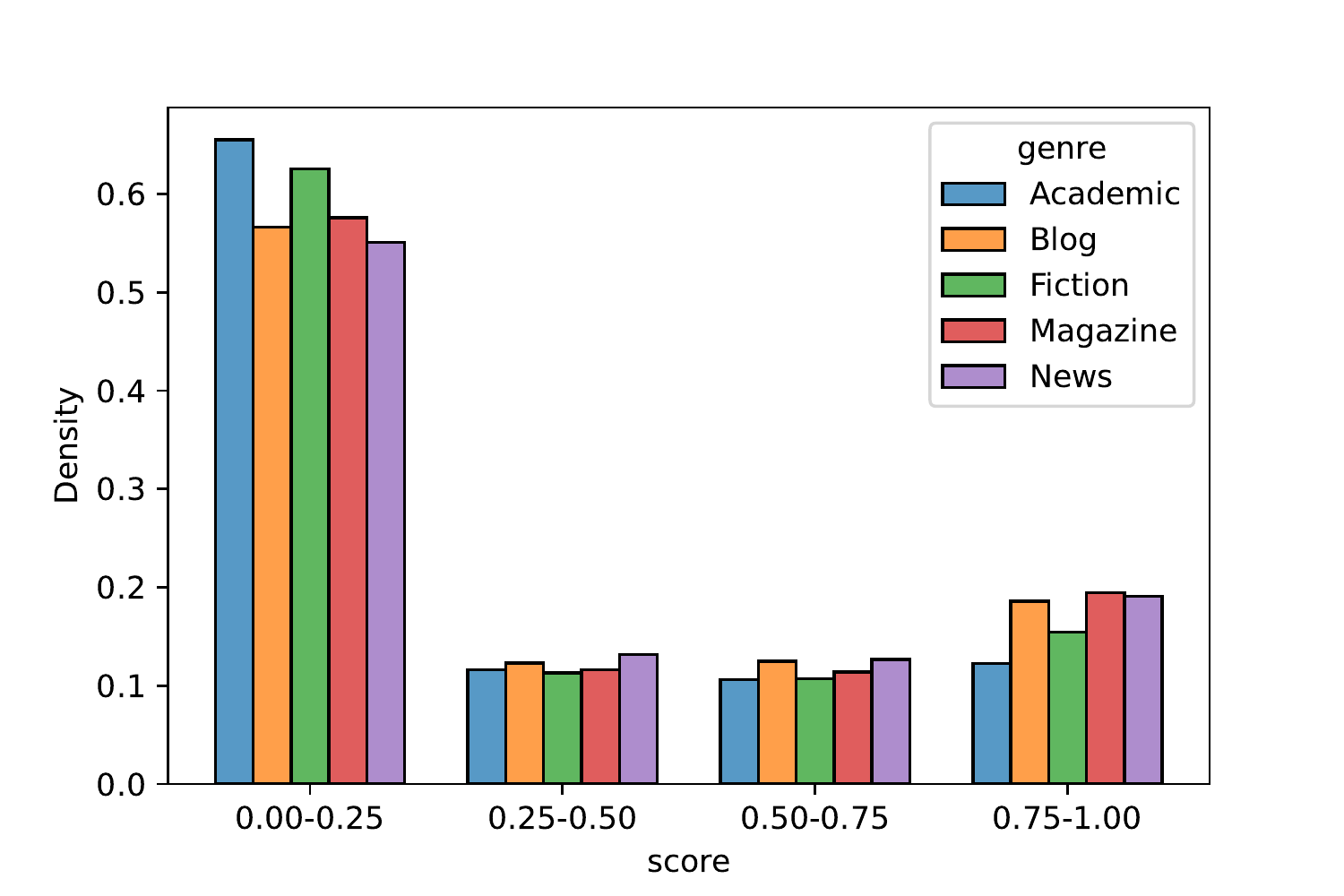}
    \vspace{-2em}
    \caption{Gap score distribution across the different genres in CoCA corpus.}\label{fig:corpus}
  \end{center}
  \vspace{-1em}
\end{figure}

\section{Quantitative Experiments}
\subsection{Settings}
To collect answers from trustworthy English speakers, we restricted the status of the workers to Masters, which means they had passed statistical monitoring and qualified as being high performers, and to those who lived in the U.S. The reward for the assignment was set to 6.5 USD, as the task was expected to take one hour and this is the average hourly payment on MTurk. We did not allow the same workers to participate in the task again.
We also briefly checked for invalid submissions, such as filling in the blanks with the same word or random words, and confirmed that all workers had completed the tasks without any significant misunderstanding.
We set the parameter $k=2$ defined in Eq.~(7).

\subsection{Corpus comparison}
 To investigate the impact of using different input data sources, we conducted additional  comparative experiments using five different genres in the CoCA corpus: \textit{academic}, \textit{blog}, \textit{fiction}, \textit{magazine}, and \textit{news}. Forty target words (as in the quantitative experiment) were used to generate 264k questions in total with \textit{roberta-base}.
 Figure~\ref{fig:corpus} shows the distribution of gap score for each genre. As we can see, the overall distribution was quite similar across all genres. However, there was a slight difference in each bin. The density for scores 0.75--1.00 was 0.12, 0.18, 0.15, 0.19, and 0.19 for the five respective genres. These scores indicate that the model was less capable of generating single answer questions for the \textit{academic} genre. This is presumably because there were less  frequent words in academic texts compared to other genres' texts. Generally, MLMs capture correlations across the words in a dataset in a data-driven manner. Therefore, the model has little knowledge about the less frequent words, and predicting a masked target with such unfamiliar words may degrade the performance. Additionally, if words are too infrequent, the tokens are substituted with an [UNK] token, and this will also affect the predictability.

\subsection{Model comparison}
The full results of the correlation plot are presented in Fig.~\ref{fig:gap_vs_corr_all}. In the highly correlated models, that is, the BERT and RoBERTa variants, a higher gap score indicated a higher correct ratio. However, the models with lower correlation, the DistilBERT and ALBERT variants, tended to under-rate samples with a high correct ratio.

We had expected roberta-large to perform better, but its correlation coefficient was lower than that of roberta-base. We found that, compared to roberta-base, roberta-large had a greater number of samples whose gap scores were high but correct ratios were low. This means that roberta-large managed to predict the target words better than humans.

\begin{figure*}[th]
\centering
  \includegraphics[clip,width=\linewidth]{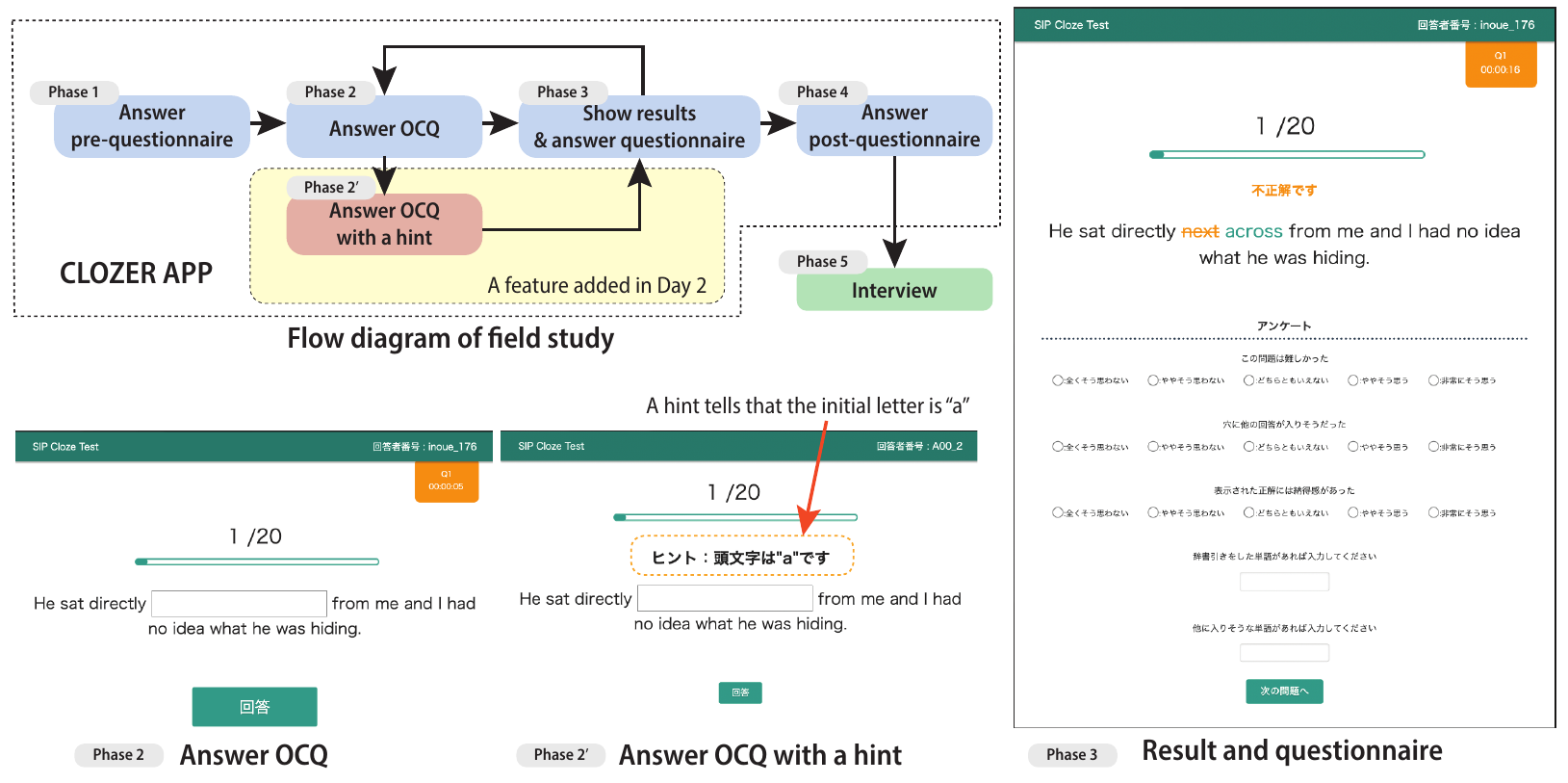}
  	\vspace{-5mm}
  	\caption{Flow diagram of the field study. The process of the field study conducted at a local high school is shown at the top left, surrounded by sample screenshots of the application. 
    }
  	\label{fig:field}
\end{figure*}
\section{Field Study}

\subsection{Aim}
We conducted a field study at a local high school to clarify the benefits and hurdles when introducing open cloze tests generated by CLOZER to the educational front. A prototype of the CLOZER application, ``CLOZER proto'', was implemented for this purpose. We deployed CLOZER proto on a web platform accessible from any device using a browser.

Our research questions for the field study were as follows.
\begin{itemize}
    \item[RQ1] What are the impressions for the CLOZER application? (Impressions)
    \item[RQ2] How appropriate are the generated questions for L2 English learners? (Appropriateness)
    \item[RQ3] What are the desirable features for the CLOZER application? (Features)
\end{itemize}
We expected most of the students to be unfamiliar with open cloze tests. Thus, in RQ1, our aim was to uncover the students' impressions of the application, especially the interest in using such an application for their learning. In RQ2, we evaluated whether that open cloze questions generated by \prop{} were suitable for L2 learning. Although we confirmed in the qualitative experiments that native speakers could correctly answer the open cloze questions generated by CLOZER, we wanted to see if the difficulty of the questions was appropriate for L2 learning high school students. Finally, in RQ3, our goal was to collect a list of features to improve CLOZER proto.

\subsection{Settings}
The field study was conducted at a high school in Kanagawa, Japan. For this study, 24 high school students in either their first or second year were recruited.
The study was conducted two days after school with different participants, who were asked to sign a consent form beforehand.

An overview of the field study is shown in Fig.~\ref{fig:field}. First, to determine the participants' ability and attitude regarding English, they were asked to fill in pre-questionnaires.
Next, they took the examination, which consisted of 20 open cloze questions generated by CLOZER. For each question, after submitting an answer, participants were asked to answer three per-question questionnaires (MidQ1--3) related to the appropriateness of the question~(see Table~\ref{tab:questionnaires}).
After submitting all answers, participants were asked to answer post-questionnaires~(PosQ1--3, Table~\ref{tab:questionnaires}) to share their impressions of the prototype.
All answers were set to a five-level Likert scale.
Finally, an open-ended interview was conducted in which participants were asked three questions regarding the difficulty of the test and potential new features we could add~(Table~\ref{tab:questionnaires}).

We followed the same strategy as the quantitative experiments to generate target questions.
Specifically, we set the level of vocabulary used in the questions so that it would be familiar to high school students by restricting the target words and target sentences with the word list.

The correct overall ratio was measured by two metrics: exact match and stem match.
The exact match regards the answer to be correct when it is identical to the ground truth, and the stem match is correct when the stem of the answer matches the stem of the ground truth.
We used Snowball stemmer~\cite{porter2001snowball} to compute the stem of each word. Both of the criteria above were case insensitive.

\begin{table*}[t]
\begin{tabular}{@{}lllcc@{}}
\toprule
Asked in                & Identifier & Question                                                   & Type           & Related RQ \\ \midrule
\multirow{2}{*}{Phase1} & PreQ1      & I like studying English.                                   & 5-point Likert & RQ1, RQ2   \\
                        & PreQ2      & I am good at English.                                      & 5-point Likert & RQ1, RQ2   \\ \midrule
\multirow{3}{*}{Every Phase3} & MidQ1    & This question was difficult.                                              & 5-point Likert & RQ2      \\
                                 & MidQ2    & There are words other than the correct answer that will fit in the blank. & 5-point Likert & RQ2      \\
                                 & MidQ3         & The presented correct answer was reasonable.               & 5-point Likert & RQ2        \\ \midrule
\multirow{3}{*}{Phase4}          & PosQ1 & The open cloze test was interesting.                                      & 5-point Likert & RQ1      \\
                        & PosQ2      & I am willing to use this application again.                & 5-point Likert & RQ1        \\
                        & PosQ3      & This application is useful for language learning.          & 5-point Likert & RQ1        \\ \midrule
\multirow{4}{*}{Interview}       & IntQ1  & How difficult was the test and why?                                       & Free form      & RQ1, RQ3 \\
                        & IntQ2       & What kind of new features do you think would be useful? & Free form      & RQ3        \\
                        & IntQ3       & Other opinions (Optional).                                  & Free form      & All        \\
                        & IntQ4       & What did you think about the hints? Were they useful?             & Free form      & RQ2, RQ3   \\ \bottomrule
\end{tabular}
\caption{Questionnaires used in the field study.
}
\label{tab:questionnaires}
\end{table*}

\begin{figure*}[t]
  \centering
  \includegraphics[width=.8\linewidth]{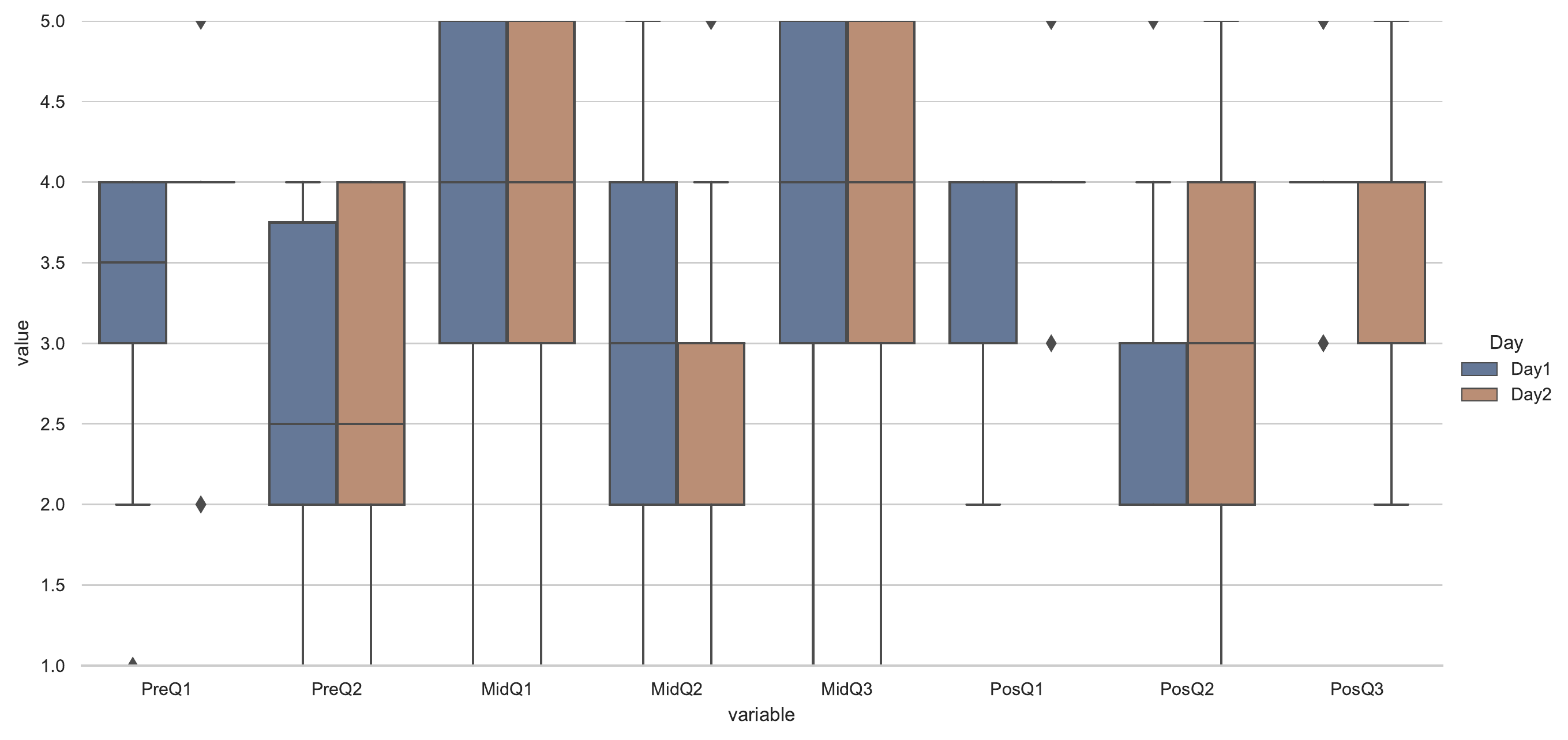}  
  \caption{Results of questionnaires in the field study.
  For the content of each questionnaire, see Table.~\ref{tab:questionnaires}.
  Note that while PreQ1--2 and PreQ1--3 were collected from each student and MidQ1--3 were from each question.
  }
  \label{fig:perquestion}
  \label{fig:preandpos}
  \label{fig:field_results}
\end{figure*}

\begin{figure*}[t]
\centering
\begin{subfigure}{.45\textwidth}
  \centering
  \includegraphics[width=\linewidth]{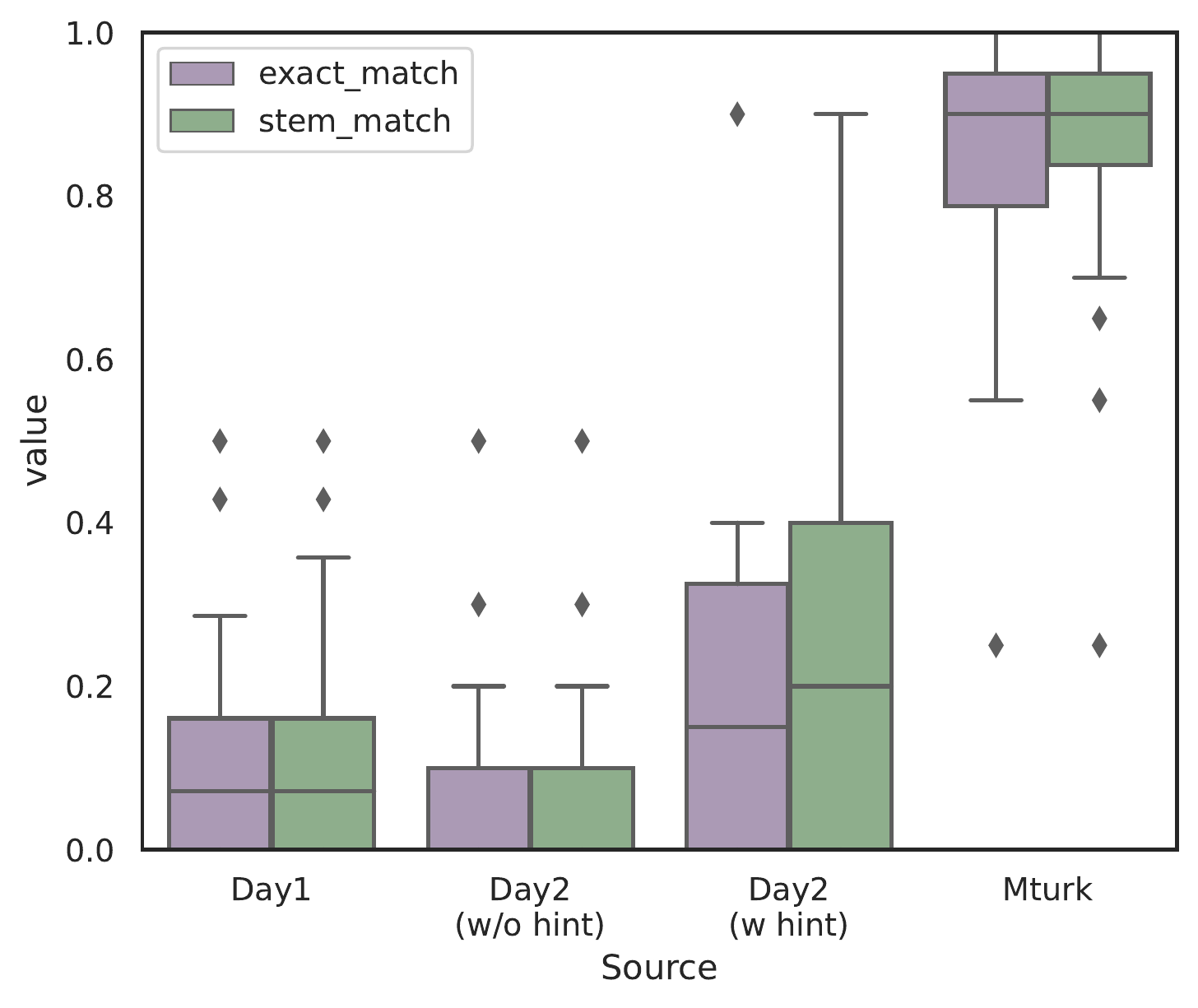}  
  \caption{Correct ratio grouped by question id.}
  \label{fig:correctr_per_qid}
\end{subfigure}
\begin{subfigure}{.45\textwidth}
  \centering
  \includegraphics[width=\linewidth]{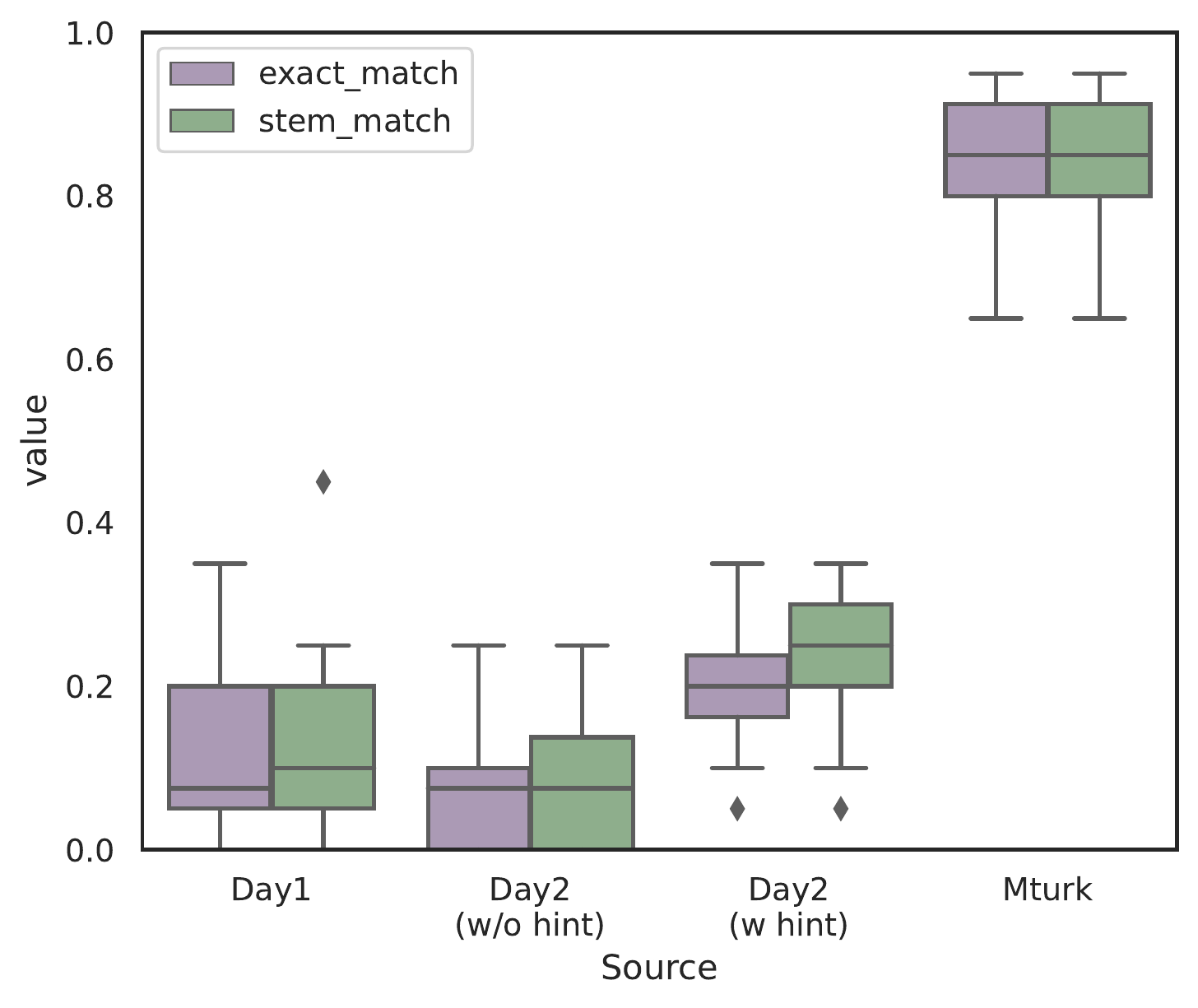}  
  \caption{Correct ratio grouped by users.}
  \label{fig:correctr_per_user}
\end{subfigure}
\caption{Correct ratio for the questions used in the field study.
The correct ratios for the high school students (day 1, day 2 without a hint, and with a hint) and for the native speakers from Amazon Mechanical Turk (MTurk) are presented.
\textit{``Exact match''} regards the answer to be correct when it is identical to the ground truth and the \textit{``stem match''} is correct when the stem of the answer matches the stem of the ground truth.
}
\label{fig:field_correctr}
\end{figure*}
\subsection{Day 1}\label{ssr:day1}
On the first day, 14 students, six in their first year and eight in their second year, participated in the study.
Among these 14 students, 11 were qualified as EIKEN Grade Pre-2, which is equivalent to the European Framework of Reference for Languages (CEFR) A2 level; two were qualified as EIKEN Grade 3, which is equivalent to the CEFR A1 level, and one was not qualified with any such tests.
The results of the pre-questionnaires (PreQ1 and PreQ2) were $3.21{\pm0.97}$ and $2.57{\pm1.16}$ respectively. Figure.~\ref{fig:preandpos} indicate that most of the students had neutral impressions about English (c.f. PreQ1), and the majority self-reported that they were not good at English (c.f. PreQ2).

The results of the per-question questionnaires (MidQ1--3) were, $3.96{\pm 1.09}$, $2.80{\pm 1.09}$, and $3.78{\pm 1.06}$ respectively (Fig.~\ref{fig:perquestion}).
The results for correct ratio are presented in Fig.~\ref{fig:field_correctr}.
The exact match ratio for users was $11.79{\pm 11.20}$ on average, where the minimum value was $0.00$ and the maximum was $35.00$.
The stem match ratio, on the other hand, was $13.21{\pm12.65}$ on average, where the minimum value was $0.00$ and the maximum was $45.00$.

Finally, the results of the post-questionnaires (PosQ1--Q3) were $3.43{\pm 0.76}$, $2.86{\pm0.86}$, and $4.07{\pm0.62}$, respectively (Fig.~\ref{fig:preandpos}).
In the interview IntQ1, most of the students reported that the test was too difficult. For example, one student said \textit{``It was difficult. I could not come up with any words.''}

\subsection{Day~2}\label{ssr:day2}
On the second day, ten students, all of whom were in their first year, participated in the study. Among these ten students, three were qualified as EIKEN Grade Pre-2 four were qualified as EIKEN Grade 3, and four was not qualified with any such tests. On the basis of feedback obtained on the first day, we decided to ease the difficulty of the test by adding a hint to each question (see Fig.~\ref{fig:field}). The hint was the first letter of the ground truth word given on each question when participants had answered incorrectly on their first try. The idea was essentially to give them a second chance for each question in this setting. To determine the effect of this new feature, we added a new question to the interview: ``What did you think about the hints? Were they useful?''

The results for the pre-questionnaires (PreQ1 and PreQ2) were $3.70{\pm 0.94}$ and $2.80{\pm 1.14}$ respectively.
Figure.~\ref{fig:preandpos} shows, compared to the first day, most of the students had a positive attitude toward English (c.f. PreQ1) and their self-reported English proficiency was slightly higher (c.f. PreQ2).
 
The results for the per-question questionnaires (MidQ1--3) are presented in Fig.~\ref{fig:perquestion}, where we can see that the averages were $3.91{\pm 1.07}$, $2.49{\pm 1.00}$, and $3.97{\pm 0.85}$, respectively. 
The results for correct ratio are presented in Fig.~\ref{fig:field_correctr}.
The exact match ratio without a hint was $8.00{\pm 8.50}$ on average, where the minimum value was $0.00$ and the maximum was $25.00$, while adding a hint increased the score significantly to $19.50$ on average, where the minimum value was $5.00$ and the maximum was $35.00$. The stem match ratio without a hint, on the other hand, was $8.50$ on average, where the minimum value was $0.00$ and the maximum was $25.00$, while adding a hint also increased the score significantly to $23.00$ on average, where the minimum value was $5.00$ and the maximum was $35.00$.

Finally, the results for the post-questionnaires (PosQ1--Q3) were $4.10{\pm 0.57}$, $3.00{\pm1.24}$, and $3.60{\pm1.07}$, respectively (Fig.~\ref{fig:preandpos}). 

\subsection{Native results}
To determine the upper bound of the correct ratio for the questions, we also collected answers from 20 native English speakers using Amazon Mechanical Turk. The conditions for the task were the same as in the qualitative experiments, except for the number of questions. The average correct ratio for the native speakers was $84.00\%$, where the maximum was $95.00\%$ and the minimum was $65.00\%$ (these results are also shown in Fig.~\ref{fig:field_correctr}).
\section{Discussion}
\subsection{Impressions for open cloze questions (RQ1)}
As shown in the Fig.~\ref{fig:preandpos}, the results for PosQ1--3, which asked about the impression of the application prototype, indicate that, despite the difficulty of the tests, most of the students found that the open cloze questions were interesting (PosQ1) and useful for language learning (PosQ3). For example, one student said in the interview that \textit{``I think this application will actually improve my English if I continue using it.''} Another said \textit{``Five questions per day will be good. I think I can make use of this application while I'm commuting to school.''} However, the results for PosQ2, which asked about the eagerness to use the application, were below average overall. The interview responses indicated that the difficulty of the questions and the lack of auxiliary functions to support learning were the leading causes of this result (see subsections \ref{sssr:difficulty} and \ref{sssr:functions}). Although the application we used in the study was only a prototype of CLOZER, limited as such to its basic functions, this finding indicates that the application will need some improvements before it is ready for actual use. This is explored in greater detail in the following subsetctions.

\subsection{Difficulty of the questions (RQ2)}\label{sssr:difficulty}
As indicated by the overall correct ratio (Fig.~\ref{fig:field_correctr}) and by the responses to MidQ1 that asked about the difficulty of the question (Fig.~\ref{fig:perquestion}), most of the questions were considered difficult by the students in this study. While most of the students were satisfied with the presented answer (c.f. Fig.~\ref{fig:perquestion} MidQ3), some of them were not convinced that it was the only correct answer. This is presumably because, since we did not show the explanations for each question, students may not have realized how suitable the true answer was. 

Most of the students were unfamiliar with open cloze tests. For example, one student said in the interview (Int1), \textit{``I usually do multiple choice questions. It was difficult without choices since I could not use the elimination method.''}

The responses to Int1 revealed there were four main types of difficulty:
\begin{itemize}
\item[D1] Words were unfamiliar overall.
\item[D2] Grammar was unfamiliar.
\item[D3] Words were familiar but student failed to grasp the context.
\item[D4] Words were familiar and context was understood, but student failed to come up with a word.
\end{itemize}

Not many students reported that words used for both questions and answers were unfamiliar (D1). Although we tried to avoid such a problem by restricting words (excluding proper nouns) to those on the vocabulary list and allowing students to use a dictionary during the test, students still found the words difficult. Most of them said that while dictionaries were helpful, they still needed to determine the meaning of a word from among several possible interpretations, which was a hurdle for some. The inclusion of proper nouns was also tricky, as these are unlikely to be in a dictionary unless they are famous.

A few students reported that the grammar was difficult (D2). For example, one student said \textit{``Usually when I do questions I know the grammar knowledge that is being tested. In this test, I was confused because there were no such instructions.''} Most of the students reported that the words were familiar, but they struggled to grasp the context (D3). For example, one student said \textit{``I was able to guess the parts of speech for the blank, but I could not find the connection between the sentence before and after the blank.''} Another student said \textit{``I know all the words, but when they come together, I wasn't able to understand what the sentence was saying.''} Similarly, most of the students reported that, even when both words and context were understandable, they often failed to come up with a word for the blank (D4). For example, one student said \textit{``I generally understood the words, but mostly I could not come up with a word for the blank.''} As for the hint feature we added on day 2, most of the students found this helpful, but some said it was not enough.

\subsection{Desirable functions (RQ3)}\label{sssr:functions}
In the interview, on day 1, most of the students asked for hints, such as showing the number of letters of the correct word, showing which Japanese words would fit, and describing the proper nouns. Additionally, with regard to the difficulty, students asked for a function to change the difficulty level. For example, one student said \textit{``I am not good at English, so it would be better if the app could change the difficulty.''}, and another said \textit{``For the easier level, it would be better to have choices for the answer.''} Another common request was to show the explanation of the question after submitting an answer, such as showing the Japanese translation of the question and related grammatical knowledge. Some students said \textit{``I wanted to know why my answer was wrong. It would be helpful to show the Japanese translation and related grammatical knowledge.''} Others said that it would be beneficial to present similar questions so as to gain more substantial knowledge of the usage of the target word.

\subsection{How to ease the difficulty?}
Our findings revealed that, even when students had similar levels of English ability~(c.f. EIKEN qualifications in subsection \ref{ssr:day1} and \ref{ssr:day2}), their levels of understanding were often different (c.f. D1--4 in subsection \ref{sssr:difficulty}).
Therefore, designing hints in accordance with the level of understanding is necessary.

For example, implementing a translation function that shows the meaning of each word will be useful for those who are less skilled at vocabulary. With this feature, students will not need to guess from among the possible word interpretations in a dictionary or consider proper nouns not included in ordinary dictionaries.

The grammatical features of each word should  also support the reading of a sentence, since L2  English learners are generally dependent on grammar rules, unlike native speakers who operate such rules unconsciously. Students often try to understand the meaning of a sentence by using their grammatical knowledge, but extracting such information can be difficult for some of them (c.f. D2, subsection \ref{sssr:difficulty}). One solution would be to show a part of speech and a dependency of each word. The part of speech (noun, verb, adjective, adverb, pronoun, preposition, conjunction, interjection, numeral, article, and determiner ) indicates how the word functions in terms of both meaning and grammar within the sentence, while the dependency shows which word is related to another word as well as the type of the relationship between them. With such information, users will be able to use their grammatical knowledge to understand the meaning of a sentence. Moreover, as a clue to the answer, showing the inflected form and the part of speech of the target word will help users to determine the grammatical aspects of the answer. For example, if users know that the answer is a verb and in the past tense, they can concentrate on finding verbs in the past tense. Most of such grammar information can be extracted using a dependency parser; specifically, the parts of speech can be obtained from POS tags, and the dependencies can be obtained from a dependency tree.

In the test, most of the students struggled to grasp the context of a sentence (c.f. D3, subsection \ref{sssr:difficulty}). This can be improved by adding some extra sentences before or after the target sentence. We only used a single sentence for each question in this study, and although the native speakers on MTurk managed to interpret contexts, one sentence was too short for L2 learners. Adding extra sentences will help them understand in what context the sentences were written. Showing the translation for such sentences will also be helpful.

Lastly, for those who understood the context but found it difficult to come up with the answer (c.f. D4, subsection \ref{sssr:difficulty}), it will be helpful to add additional information to the target word: for example, to show the first letter. As we found on day 2 of the field study, this hint was effective for most students. We could also increase the number of letters shown to ease the difficulty further. Similarly, showing other types of information about the target word, such as the number of letters, would also help decrease  the difficulty.

\subsection{Limitations and future studies}
While the effectiveness and benefits of CLOZER have clearly been demonstrated, there were several limitations to this study. The first lies in the masked language model utilized in the CLOZER architecture. Since this model is trained on a large corpus in a data-driven manner, it shows a better prediction if the input sentence is similar to the corpus, while in contrast, it tends to degrade the performance when the input sentence is outside the distribution of the corpus. How the corpus will affect the results is still unclear, and we leave this to future study. Additionally, as reported in our qualitative study, the model can be overconfident when the answer is not so certain. This mostly stems from the fact that the confidence values do not strictly show the prediction's certainty due to the loss function settings in the learning. However, in most cases, we found the confidence values helpful; moreover, improving the language models was outside the scope of the current study.

Additionally, for our prototype test, we only examined CLOZER on \textit{English} learning. However, our framework is theoretically not limited to a single language, as the masked language models as well as the gap score are also applicable to other languages and, in fact, there are various pre-trained models available. Extensive experiments on other languages will be desirable in future work.

Finally, since our aim in this study was to present the initial design of the CLOZER application, we did not investigate whether it is actually effective for language learning. To this end, a long-term field study to evaluate the improvement in students' English skills will be required for future research. Moreover, an examination of the application's usability for teachers will be beneficial towards the actual deployment of the application to the educational front.

{\footnotesize
\bibliographystyle{named}
\bibliography{ijcai22}
}
\end{document}